\documentclass[12pt]{article}
\usepackage{amssymb}
\usepackage{amsmath}
\usepackage{apacite}
\usepackage{latexsym}
\usepackage{epsfig}
\usepackage{graphicx}
\usepackage{subfigure}
\usepackage{wasysym}
\usepackage{threeparttable}
\usepackage{natbib}
\usepackage{color}
\usepackage{epstopdf}
\usepackage{bm}
\usepackage{float}
\usepackage{todonotes}
\usepackage{verbatim}
\usepackage{booktabs}
\usepackage{soul}

\usepackage{hyperref}

\def\log{\hbox{log}}

\def\boxit#1{\vbox{\hrule\hbox{\vrule\kern6pt
          \vbox{\kern6pt#1\kern6pt}\kern6pt\vrule}\hrule}}

\def\bse{\begin{eqnarray*}}
\def\ese{\end{eqnarray*}}
\def\be{\begin{eqnarray}}
\def\ee{\end{eqnarray}}
\def\bq{\begin{equation}}
\def\eq{\end{equation}}
\def\bse{\begin{eqnarray*}}
\def\ese{\end{eqnarray*}}

\begin{document}

\thispagestyle{empty} 
\baselineskip=28pt

\begin{center}
{\LARGE{\bf A Heterogeneous Spatial Model for Soil Carbon Mapping of the Contiguous United States Using VNIR Spectra}}

\end{center}

\baselineskip=12pt

\vskip 2mm
\begin{center}

Paul A. Parker\footnote{(\baselineskip=10pt to whom correspondence should be
  addressed) Department of Statistics, University of California, Santa Cruz, 1156 High St.,
  Santa Cruz, CA 95064, paulparker@ucsc.edu}\, and
Bruno Sansó\footnote{\baselineskip=10pt Department of Statistics, University of California, Santa Cruz, 1156 High St.,
  Santa Cruz, CA 95064,
  bsanso@ucsc.edu}

\end{center}

\vskip 4mm
\baselineskip=12pt 
\begin{center}
{\bf Abstract}
\end{center}

The Rapid Carbon Assessment, conducted by the U.S. Department of Agriculture, was implemented in order to obtain a representative sample of soil organic carbon across the contiguous United States. In conjunction with a statistical model, the dataset allows for mapping of soil carbon prediction across the U.S., however there are two primary challenges to such an effort. First, there exists a large degree of heterogeneity in the data, whereby both the first and second moments of the data generating process seem to vary both spatially and for different land-use categories. Second, the majority of the sampled locations do not actually have lab measured values for soil organic carbon. Rather, visible and near-infrared (VNIR) spectra were measured at most locations, which act as a proxy to help predict carbon content. Thus, we develop a heterogeneous model to analyze this data that allows both the mean and the variance to vary as a function of space as well as land-use category, while incorporating VNIR spectra as covariates. After a cross-validation study that establishes the effectiveness of the model, we construct a complete map of soil organic carbon for the contiguous U.S. along with uncertainty quantification.

\baselineskip=12pt
\par\vfill\noindent
\par\medskip\noindent
\clearpage\pagebreak\newpage \pagenumbering{arabic}
\baselineskip=24pt

\section{Introduction}

Soil organic carbon (SOC) content is an important measurement of soil quality and health \citep{nunes2021soil}. Furthermore, carbon sequestration has important implications in regards to climate change \citep{smith2020measure}. Thus, understanding the spatial distribution of SOC constitutes an important problem. However, collection of soil carbon data tends to be very localized, limiting the ability to model SOC at the continental scale.

The Rapid Carbon Assessment (RaCA) was conducted by the Natural Resource Conservation Service in order to better understand the distribution of soil carbon across the conterminous U.S. (CONUS), through a large-scale data collection effort that spanned both regions as well as land-use categories \citep{wills2014overview}. The RaCA resulted in data collected at roughly 32,000 unique locations over roughly 6,000 field sites. However, due to the cost of measuring SOC in a lab, SOC was not directly measured for most soil samples. Instead, a visible and near-infrared spectrometer (VNIR) was used on the samples, resulting in reflectance spectra from 350-2,500 nm. These spectra, which are highly correlated with SOC, were then used to predict SOC for most soil samples.

Our goal is to use the RaCA data to predict SOC at the surface level across the CONUS. In doing so, we are mindful of the fact that most soil samples do not contain lab measured values of SOC, but rather predictions. This motivates the need for a model that uses the spectral data as covariates to predict SOC. Furthermore, exploratory analysis reveals that the central tendency as well as the dispersion of SOC seem to vary spatially as well as by land-use category, motivating the need for a model that can handle highly heterogeneous data. Ultimately, our proposed model allows the conditional mean and variance both to vary across space and by land-use category, while considering the VNIR spectra as covariates. Lastly, as the RaCA contains many data points, we rely on recent distribution theory related to Bayesian modeling for heteroscedastic data in order to fit our proposed model in a computationally efficient manner.

To date, a number of studies have utilized the RaCA, although not for the purpose of predicting SOC across the CONUS, while acknowledging the uncertainty attributable to the spectra-based predictions. \cite{wijewardane2016prediction} study a variety of models that may be used to predict SOC based on VNIR spectra. However, their approach is strictly for predicting SOC at locations where VNIR spectral data are sampled. Thus, these methods are not readily equipped to predict SOC across the CONUS. \cite{risser2019nonstationary} focus on the closely related problem of predicting soil carbon stocks using the RaCA, using various nonstationary spatial models. However, an important distinction from our goal is that they do not consider the VNIR spectra in their approach. Another distinction is that they use covariate driven partitioning (i.e. land-use category), to in a sense predict the covariate values for out of sample locations. In contrast to this, we obtain land-use category at a high resolution and treat the covariates as known. Lastly, they work with a subregion of the CONUS, whereas we consider the entire CONUS. At this larger scale, the need for a more complex variance model is apparent.

{ There exists some related literature on statistical modeling with the use of spectroscopic covariates. \cite{brown2001bayesian} consider an expansion of wavelet basis functions for near infrared spectra, and use this to predict composition of bread dough. A Bayesian variable selection approach is used to identify the basis functions most closely associated with the response. \cite{chakraborty2012bayesian} consider the use of a Bayesian support vector regression, again to predict dough composition based on spectral information. Finally, \cite{stingo2012bayesian} and \cite{gutierrez2014bayesian} construct classification models to determine the source of different meats based on spectral inputs. However, these approaches are all intended for independent observations and are thus not equipped to handle spatial correlation. In contrast to this, \cite{yang2015bayesian} do consider spatial correlation in order to predict soil electrical conductivity by depth profiles for 26 sites in Missouri.

The literature on modeling of spatially non-stationary random fields is also closely related to our work. Much of the early work in this area is based on the idea of deforming the spatial domain to a latent space where stationarity holds (e.g. see \cite{sampson1992nonparametric} and \cite{schmidt2003bayesian}). More recently, \cite{zammit2022deep} use deep learning to estimate the deformation function. Another common approach involves partitioning the domain into a set of locally stationary processes \citep{gramacy2008bayesian, kim2005analyzing, risser2019nonstationary}. Process convolutions may also be used by allowing for a kernel that evolves over space \cite{higdon1998process, lemos2009spatio}. Along these lines, \cite{kirsner2020multi} consider a multi-resolution kernel model. Finally another method involves a hierarchical model where a spatially informed Inverse Wishart prior distribution is used for the covariance matrix \citep{brown1994multivariate, GreSanMatt2023}. For other examples of nonstationary spatial covariance structure, see \cite{schmidt2020flexible} and the references therein.
}

The remainder of this work is outlined as follows. Section~\ref{sec: data} describes the RaCA data in further detail and provides some brief exploratory analysis. Following this, we present our proposed model and relevant background knowledge in Section~\ref{sec: methods}. Section~\ref{sec: cv} provides a cross-validation study that is used to guide the model choice for carbon prediction, while Section~\ref{sec: analysis} utilizes the selected model and the RaCA data to produce predictions as well as uncertainty estimates of SOC for the CONUS. Finally, we provide some discussion and concluding remarks in Section~\ref{sec: discussion}.

\section{RaCA Data}\label{sec: data}

\begin{figure}[H]
\centering
\includegraphics[width=150mm,height=80mm]{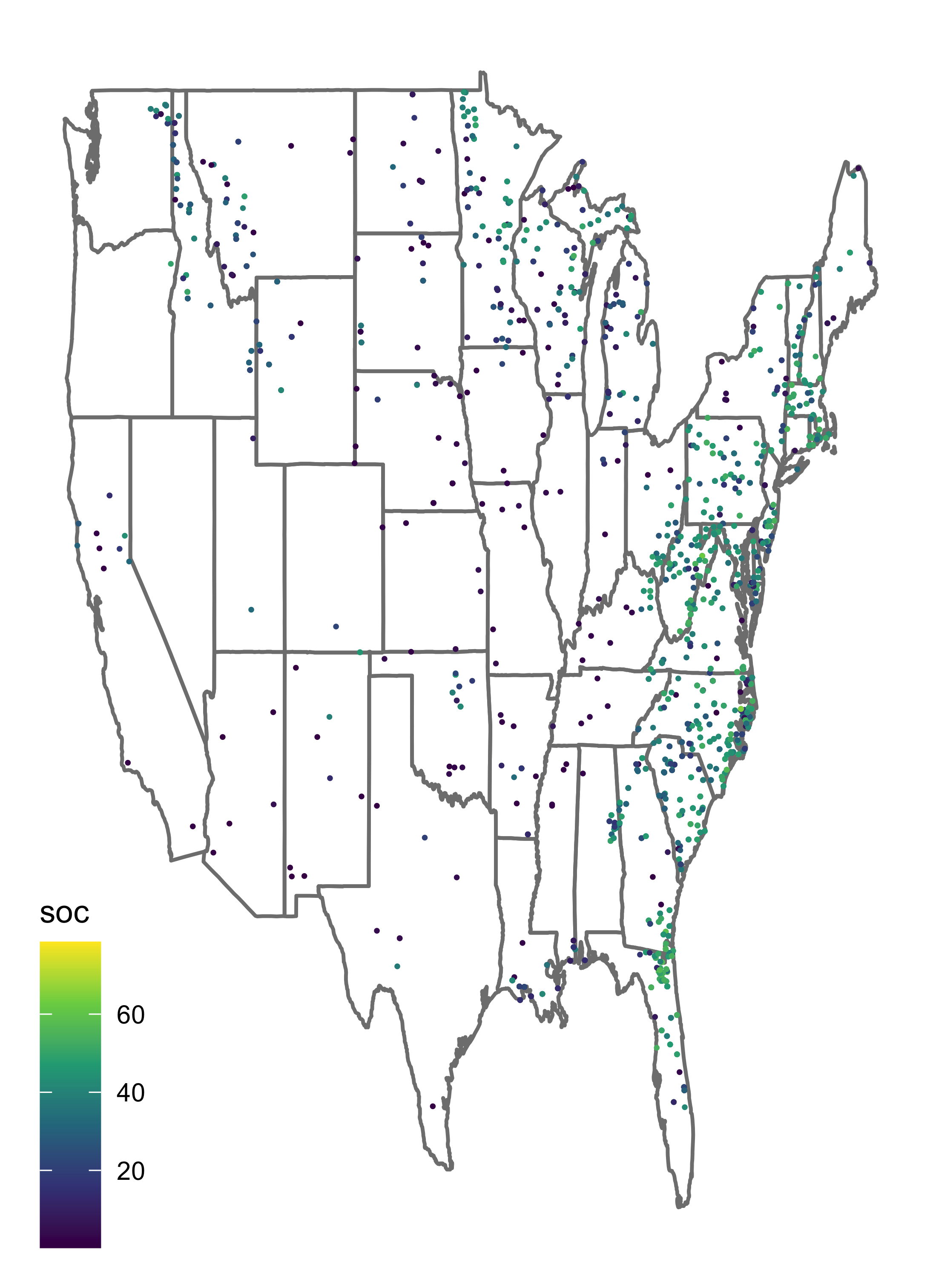}
\caption{Locations and soil organic carbon measurements for sites with lab data.} 
\label{FIG: soc}
\end{figure}

\begin{figure}[H]
\centering
\includegraphics[width=150mm,height=80mm]{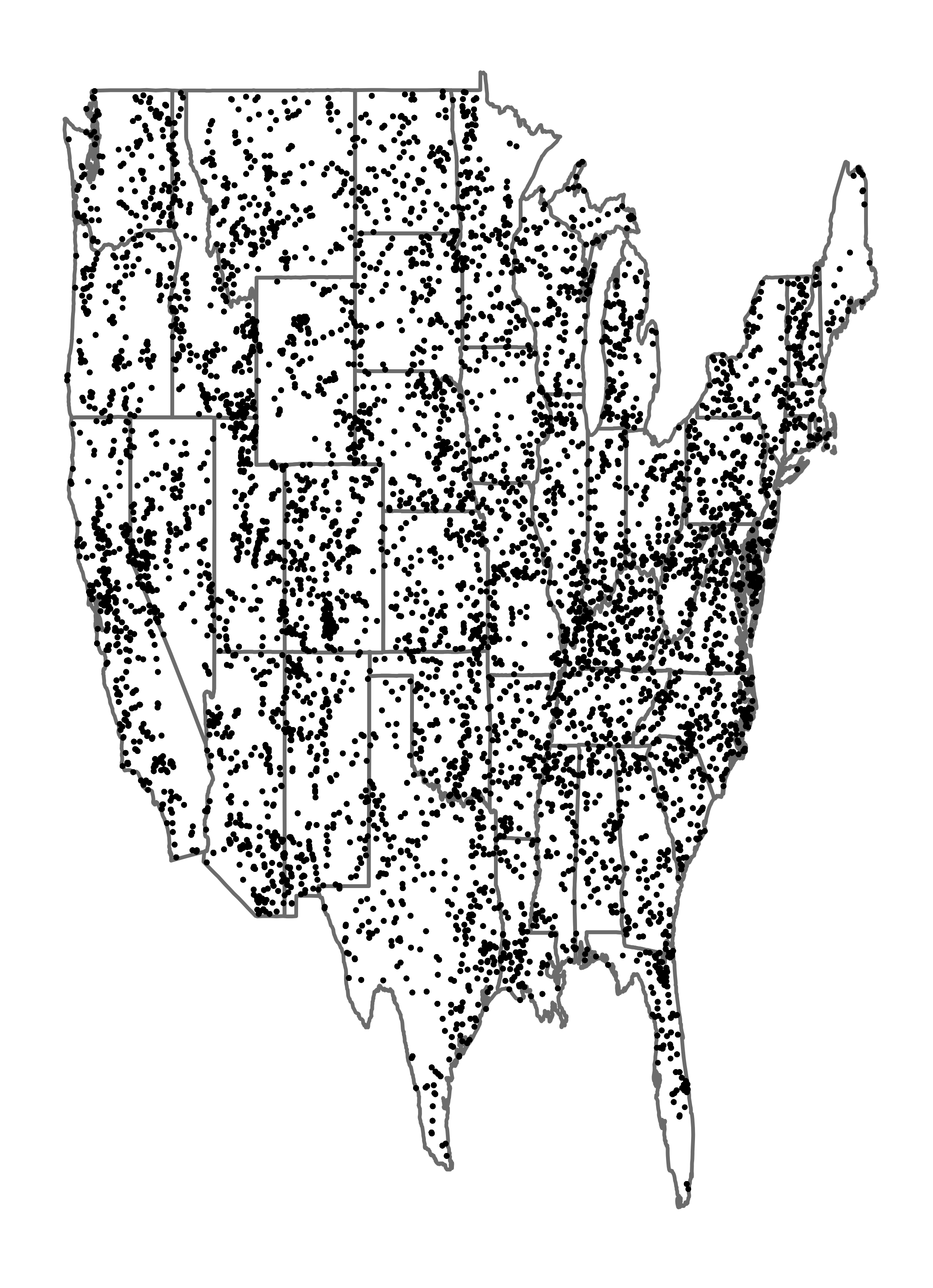}
\caption{Locations of all RaCA sites with VNIR spectra.} 
\label{FIG: all}
\end{figure}

\begin{figure}[H]
\centering
\includegraphics[width=150mm,height=80mm]{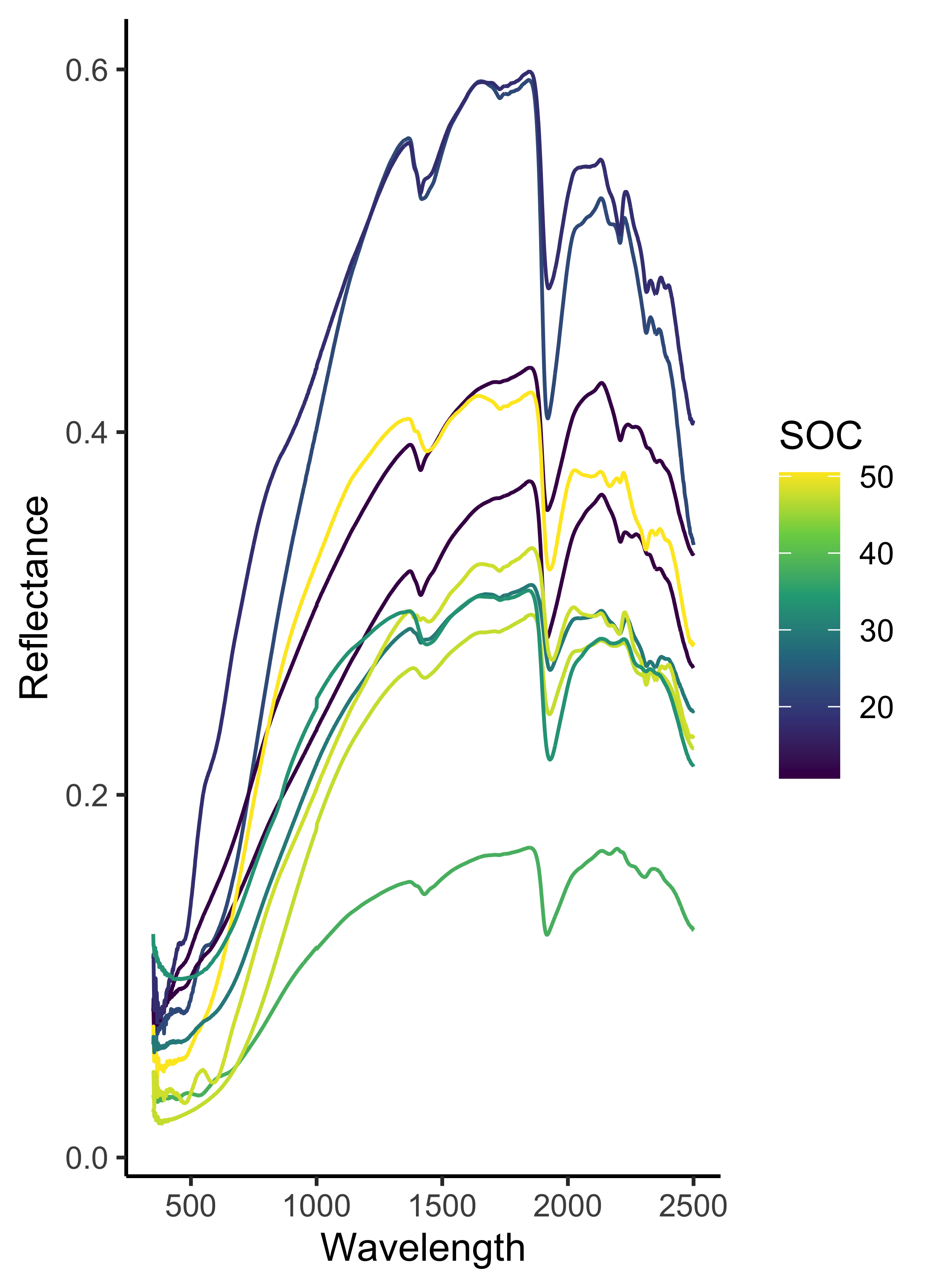}
\caption{VNIR spectra for ten randomly selected soil samples in the Rapid Carbon Analysis, along with soil organic carbon measures.} 
\label{FIG: spec}
\end{figure}

Herein, we consider data from the Rapid Carbon Assessment (RaCA). Data were collected at a variety of depths, but  we only consider data collected at the surface level. In total, there are 28,010 locations with data, denoted by $\bm{s} \in \mathcal{S}.$ Of these, only 3,093 sites have lab measurements of soil organic carbon.  We denote these sites by $\mathcal{S}_{\hbox{lab}} \subset \mathcal{S},$ and denote the log-transformed soil carbon measurements as $y(\bm{s}).$ VNIR spectra are measured for all $\bm{s} \in \mathcal{S}.$ VNIR data consists of reflectance measurements $r(\omega, \bm{s})$ for wavelengths $\omega=350, 351,\ldots, 2500$ nm and sites $\bm{s} \in \mathcal{S}.$ A map of soil carbon measurements for $ \bm{s} \in \mathcal{S}_{\hbox{lab}}$ is given in Figure~\ref{FIG: soc}. A map of all $\bm{s} \in \mathcal{S}$ is given in Figure~\ref{FIG: all}. {Finally, Figure~\ref{FIG: spec} shows the VNIR spectra for ten arbitrary sites in the RaCA data, along with their measure of SOC. Visually, there does seem to be a relationship between the spectra and SOC for this limited subsample.}

\begin{figure}[H]
\centering
\includegraphics[width=150mm]{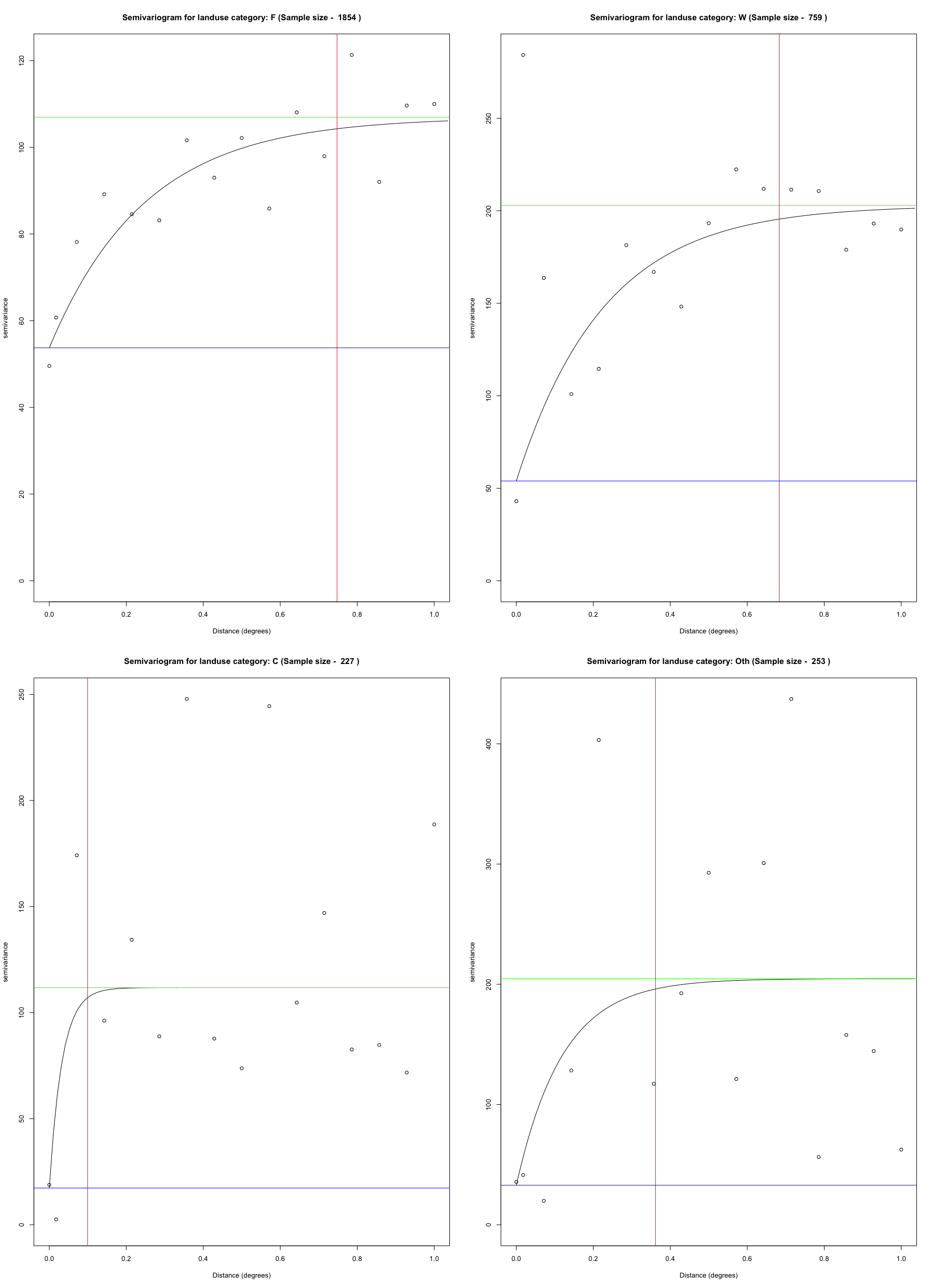}
\caption{Semivariograms for each land-use category.} 
\label{FIG: svg}
\end{figure}

In addition to soil carbon measurements, soil samples have an associated land-use category. These are classified as cropland (C), forestland (F), wetland (W), or other (Oth). As an exploratory analysis to assess spatial dependence, Figure~\ref{FIG: svg} shows the semivariograms for each land-use category. There is clear heterogeneity across the soil land-use categories, whereby both the range of spatial dependence as well as the scale of variation appear to differ across the different categories. This indicates that the land-use category plays an important role in soil carbon sequestration, and motivates the need for a flexible model that allows for the distributional assumptions to vary by land-use category.

\section{Methodology}\label{sec: methods}

We consider a variety of modeling approaches to predict soil carbon across the CONUS. Based on the exploratory analysis presented in Section~\ref{sec: data}, we focus on heteroscedastic models. In other words, we want to allow both the mean function and the variance function for soil carbon to vary across land-use category as well as spatially, with potential interactions between the two.

To allow for flexible modeling of the variance function, we utilize ideas presented by \cite{parker2021general}. They present a general Bayesian model for heteroscedastic data that utilizes a conjugate prior framework allowing for efficient sampling from the posterior distribution. Specifically, the framework relies on the recently developed multivariate log-gamma distribution (MLG). The MLG was introduced by \cite{bradley2018computationally} and \cite{bradley2019bayesian}, with the original purpose of acting as a conjugate prior in Poisson regression. The density for the MLG distribution is given as 
\begin{equation*}
    \hbox{det}(\bm{VV'})^{-1/2}
    \left\{ \prod_{i=1}^n \frac{\kappa_i^{\alpha_i}}{\Gamma(\alpha_i)}\right\}
    \hbox{exp}\left[\bm{\alpha}' \bm{V}^{-1}(\bm{Y- \mu}) -
    \bm{\kappa}' \hbox{exp}\left\{\bm{V}^{-1}(\bm{Y- \mu}) \right\}\right],
\end{equation*} and denoted by $\hbox{MLG}(\bm{\mu}, \mathbf{V}, \bm{\alpha}, \bm{\kappa})$. The parameters consist of a length $n$ centrality vector, $\bm{\mu},$ an $n\times n$ matrix $\bm{V}$ that controls the correlation structure, as well as length $n$ shape and rate vectors, $\bm{\alpha}$ and $\bm{\kappa}.$

One can easily sample a length $n$ vector $\bm{Y}\sim \hbox{MLG}(\bm{\mu}, \mathbf{V}, \bm{\alpha}, \bm{\kappa})$ through the following steps:
\begin{enumerate}
        \item Generate a vector $\mathbf{g}$ as $n$ independent Gamma random variables with shape $\alpha_i$ and rate $\kappa_i$, for $i=1,\ldots,n$
        \item Let $\mathbf{g}^*=\hbox{log}(\mathbf{g})$
        \item Let $\mathbf{Y}=\mathbf{V g}^* + \bm{\mu}$.
    \end{enumerate} Additionally, Bayesian inference with MLG priors will require simulation from the conditional multivariate log-Gamma distribution ($\hbox{cMLG}$). Letting $\mathbf{Y} \sim \hbox{MLG}(\bm{\mu},  \mathbf{V}, \bm{\alpha}, \bm{\kappa})$, \cite{bradley2018computationally} show that $\mathbf{Y}$ can be partitioned into $(\mathbf{Y_1}', \mathbf{Y_2}')'$, where $\mathbf{Y_1}$ is $r$-dimensional and $\mathbf{Y_2}$ is $(n-r)$-dimensional. The matrix $\mathbf{V}^{-1}$ is also partitioned into $\left[\mathbf{H \; B}  \right]$, where $\mathbf{H}$ is an $n \times r$ matrix and $\mathbf{B}$ is an $ n \times (n - r)$ matrix. Then 
$$\bm{Y_1} | \bm{Y_2} = \bm{d}, \bm{\mu}^*, \bm{H}, \bm{\alpha}, \bm{\kappa} \sim \hbox{cMLG}(\bm{\mu}^*, \bm{H}, \bm{\alpha}, \bm{\kappa})$$ with density
\begin{equation}
    M \hbox{exp} \left\{\bm{\alpha}' \bm{H Y_1} 
     - \bm{\kappa}' \hbox{exp}(\bm{H Y_1} - \bm{\mu}^*)\right\} ,
\end{equation} where $\bm{\mu}^*=\mathbf{V}^{-1}\bm{\mu} - \mathbf{Bd}$, and $\mathit{M}$ is a normalizing constant.  Importantly, \cite{bradley2018computationally} show that it is also easy to sample from the cMLG distribution using $(\mathbf{H}'\mathbf{H})^{-1}\mathbf{H}'\mathbf{Y}$, where $\mathbf{Y}$ is sampled from $\hbox{MLG}(\bm{\mu}, \mathbf{I}, \bm{\alpha}, \bm{\kappa})$.

To tackle the problem of modeling the heterogeneity of the field of soil organic carbon data we adapt the heteroscedastic model in \cite{parker2021general}.  We leverage the use of the MLG prior distribution for parameters related to the variance, in order to model that non-stationarity of the field. We formulate the model as
\begin{equation}\label{eq: mod}
    \begin{aligned}
        y(\bm{s}) | \mu(\bm{s}), \sigma^2(\bm{s}) & \stackrel{ind.}{\sim} \hbox{N}(\mu(\bm{s}), \sigma^2(\bm{s})), \; \bm{s} \in \mathcal{S}_{lab} \\
        \mu(\bm{s}) &= \bm{x}_{1}(\bm{s})'\bm{\beta}_1 + \bm{\psi}_{1}(\bm{s})'\bm{\eta}_1 \\
        -\hbox{log}(\sigma^2(\bm{s})) &= \bm{x}_{2}(\bm{s})'\bm{\beta}_2 + \bm{\psi}_{2}(\bm{s})'\bm{\eta}_2 \\
        \bm{\eta}_1 & \sim \hbox{N}(\bm{0}, \sigma^2_{\eta_1}\bm{\hbox{I}}) \\
        \bm{\eta}_2 & \sim \hbox{MLG}(\bm{0}, \alpha^{1/2} \sigma^2_{\eta_2} \bm{\hbox{I}}, \alpha \bm{1}, \alpha\bm{1}) \\
        \bm{\beta}_1 & \sim \hbox{N}(\bm{0}, \sigma^2_{\beta_1}\bm{\hbox{I}}) \\
        \bm{\beta}_2 & \sim \hbox{MLG}(\bm{0}, \alpha^{1/2} \sigma^2_{\beta_2} \bm{\hbox{I}}, \alpha \bm{1}, \alpha\bm{1}) .
    \end{aligned}
\end{equation} 
Under this approach, the spatial field is assumed to have a spatially varying mean $\mu(\bm{s})$, and variance $\sigma^2(\bm{s})   $. Similar to traditional mixed modeling approaches, the mean is modeled as a linear combination of spatially varying $p_1$-dimensional covariates  ($\bm{x}_{1}(\bm{s})'\bm{\beta}_1$) as well as a spatial random effects component ($\bm{\psi}_{1}(\bm{s})'\bm{\eta}_1$). Here,  $\bm{\psi}_{1}(\bm{s})$ is an $r_1$-dimensional vector of basis functions. Next, using a negative log link function, the variance is also modeled as a linear combination of covariates and random effects, representing spatially varying fixed and random effects respectively. Note that the variance is modeled using $p_2$-dimensional vector $\bm{x}_{2}(\bm{s})$ and $r_2$-dimensional vector $\bm{\psi}_{2}(\bm{s}),$ which may or may not be the same as the covariates and basis functions used in the model for the mean.

The use of the negative log link function along with MLG prior distributions for $\bm{\beta}_2$ and $\bm{\eta}_2$ results in straightforward sampling of cMLG full-conditional distributions for $\bm{\beta}_2$ and $\bm{\eta}_2.$  Furthermore, the specific form of MLG prior used here, namely $\hbox{MLG}(\mathbf{c}, \alpha^{1/2}\mathbf{V}, \alpha \mathbf{1}, \alpha \mathbf{1})$, is known to converge in distribution to $\hbox{N}(\bm{c}, \bm{VV'})$ as the value of $\alpha$ approaches infinity \citep{bradley2018computationally}. This allows us to harness the computational convenience of the MLG prior while selecting a prior that is effectively shaped like the normal distribution that is typically used in hierarchical modeling. { In practice we have found that there is no discernible change in the shape of the prior distribution when increasing $\alpha$ beyond 1000, and thus we use $\alpha=1000$.}

{ The model given by} (\ref{eq: mod}){ is completed by placing a prior distribution over the parameters $\sigma^2_{\eta_1}$ and $\sigma^2_{\eta_2}$. In particular, we use a conjugate Inverse Gamma prior for $\sigma^2_{\eta_1}$ with shape parameter $a$ and scale parameter $b$. We also use a conjugate log-Gamma prior truncated below at 0 for $\frac{1}{\sigma_{\eta_2}}$, with shape parameter $w$ and rate parameter $p$. In our case, we establish vague prior distributions by setting the hyperparameters $\sigma^2_{\beta_1} = \sigma^2_{\beta_2} = w = p = 1000$ and $a=b=0.5$. For other use cases, where prior information is known, the hyperparameters could be adjusted accordingly. Finally, because the model is fully conjugate, we use Gibbs sampling to sample from the posterior distribution of our model parameters. Details of the sampling approach, including full-conditional distributions are given in Appendix A.}

\section{Model selection}\label{sec: cv}

{In order to develop an approach that allows for accurate prediction of soil carbon, we consider a variety of models with increasing complexity. Then, we compare the various approaches through a cross-validation study.}

\subsection{Spatial Only Models}

As a baseline, we fit a standard linear regression

\textbf{Model 1}\begin{align*}
    y(\bm{s}) &= \gamma_{\ell(\bm{s})} + \bm{x}'(\bm{s}) \bm{\beta} + \epsilon(\bm{s}), \; \bm{s} \in \mathcal{S}_{lab} \\
    \epsilon(\bm{s}) &\stackrel{iid}{\sim} N(0, \tau^2),
\end{align*} where $\gamma_{\ell(\bm{s})}$ is a land-use specific intercept and the vector of covariates, $\bm{x}(\bm{s})$, includes longitude and latitude. This model acts as a baseline and does not include any spatial component beyond the covariates.

{Next we consider a model with explicit spatial structure, }

\textbf{Model 2}

\begin{align*}
    y(\bm{s}) &= \gamma_{\ell(\bm{s})} +  \bm{x}'(\bm{s}) \bm{\beta}_1 + \xi(\bm{s}) + \epsilon(\bm{s}), \; \bm{s} \in \mathcal{S}_{lab} \\
    \xi(\bm{s}) &= \sum_{j=1}^J \phi_j(\bm{s}) \eta_j \\
    \epsilon(\bm{s}) &\stackrel{ind}{\sim} N(0, \tau^2(\bm{s})) \\
    -\log(\tau^2(\bm{s})) &= \zeta_{\ell(\bm{s})} +  \bm{x}'(\bm{s}) \bm{\beta}_2 + \delta(\bm{s})  \\
    \delta(\bm{s}) &= \sum_{j=1}^J \phi_j(\bm{s}) \alpha_j
\end{align*} where $\phi_j(\cdot), \; j=1, \ldots,J$ are bisquare basis functions calculated using the  \texttt{FRK} pacakge. {Specifically,
$$
\phi_j(\bm{s}) = \left(1-\left(\frac{||\bm{s} - \bm{u}_j||}{R_j}\right)^2 \right)^2 I(||\bm{s} - \bm{u}_j|| < R_j),
$$ where $\bm{u}_j$ is a knot location and $R_j$ is a range parameter.} Two resolutions of basis functions are used, with 76 total basis functions. Model 2 explicitly uses both a spatially varying mean and spatially varying variance (nugget).

{Now we explore a model that includes a spatial interaction with land-use category in the mean structure,}

\textbf{Model 3}

\begin{align*}
    y(\bm{s}) &=\gamma_{\ell(\bm{s})} +  \bm{x}'(\bm{s}) \bm{\beta}_1 + \xi(\bm{s}) + \mu_{\ell}(\bm{s}) + \epsilon(\bm{s}), \; \bm{s} \in \mathcal{S}_{lab} \\
    \xi(\bm{s}) &= \sum_{j=1}^J \phi_j(\bm{s}) \eta_j \\
    \mu_{\ell}(\bm{s}) &= \sum_{j=1}^J \phi_j(\bm{s}) \kappa_{\ell j} \\
    \epsilon(\bm{s}) &\stackrel{ind}{\sim} N(0, \tau^2(\bm{s})) \\
    -\log\left(\tau^2(\bm{s})\right) &=\zeta_{\ell(\bm{s})} +  \bm{x}'(\bm{s}) \bm{\beta}_2 + \delta(\bm{s}) \\
    \delta(\bm{s}) &= \sum_{j=1}^J \phi_j(\bm{s}) \alpha_j
\end{align*} where now the process $\mu_{\ell}(\bm{s})$ is indexed by land-use category $\ell.$ Model 3 includes spatially varying mean specific to each land use category, and a spatially varying variance.

{Finally, we include a model that has a spatial interaction with land-use for both the mean and variance structure,}

\textbf{Model 4}

\begin{align*}
    y(\bm{s}) &=\gamma_{\ell(\bm{s})} +  \bm{x}'(\bm{s}) \bm{\beta}_1 + \xi(\bm{s}) + \mu_{\ell}(\bm{s}) + \epsilon(\bm{s}), \; \bm{s} \in \mathcal{S}_{lab} \\
    \xi(\bm{s}) &= \sum_{j=1}^J \phi_j(\bm{s}) \eta_j \\
    \mu_{\ell}(\bm{s}) &= \sum_{j=1}^J \phi_j(\bm{s}) \kappa_{\ell j} \\
    \epsilon(\bm{s}) &\stackrel{ind}{\sim} N(0, \tau^2(\bm{s})) \\
    -\log\left(\tau^2(\bm{s})\right) &=\zeta_{\ell(\bm{s})} +  \bm{x}'(\bm{s}) \bm{\beta}_2 + \delta(\bm{s}) + \lambda_{\ell}(\bm{s}) \\
    \delta(\bm{s}) &= \sum_{j=1}^J \phi_j(\bm{s}) \alpha_j \\ 
    \lambda_{\ell}(\bm{s}) &= \sum_{j=1}^J \phi_j(\bm{s}) \nu_{\ell j}
\end{align*}  where now the model includes both spatially varying mean and variance specific to each land-use category.

\subsection{Spectral Models}

{Models 1-4  exhibit increasingly complex spatial structure, however they do not consider the VNIR spectra, which is an important component of soil carbon prediction. To handle this, we consider a basis expansion of the spectra,
$$
r(\omega, \bm{s}) = \sum_{k=1}^K b_j(\omega) \upsilon_k(\bm{s}) + \theta(\omega, \bm{s})
$$ where $\theta(\omega, \bm{s})$ is a white noise term. Then, to include the spectra within our model, we regress on the spatially varying coefficients, $\upsilon_k(\bm{s})$. Herein, we consider the first 9 principal components of the VNIR spectra for our basis expansion, which represents over 99\% of the total spectral variation.}

{The next model we consider uses the VNIR spectra as well as a spatial interaction with land-use in the mean structure}

\textbf{Model 5}
\begin{align*}
    y(\bm{s}) &=\gamma_{\ell(\bm{s})} + \sum_{k=1}^K\upsilon_k(\bm{s})\chi_k  + \bm{x}'(\bm{s}) \bm{\beta}_1 + \xi(\bm{s}) + \mu_{\ell}(\bm{s}) + \epsilon(\bm{s}), \; \bm{s} \in \mathcal{S}_{lab} \\
    \xi(\bm{s}) &= \sum_{j=1}^J \phi_j(\bm{s}) \eta_j \\
    \mu_{\ell}(\bm{s}) &= \sum_{j=1}^J \phi_j(\bm{s}) \kappa_{\ell j} \\
    \epsilon(\bm{s}) &\stackrel{ind}{\sim} N(0, \tau^2(\bm{s})) \\
    -\log\left(\tau^2(\bm{s})\right) &=\zeta_{\ell(\bm{s})} +  \bm{x}'(\bm{s}) \bm{\beta}_2 + \delta(\bm{s}) \\
    \delta(\bm{s}) &= \sum_{j=1}^J \phi_j(\bm{s}) \alpha_j.
\end{align*} Model 5 is identical to model 3, however, now the mean is a linear function of the spectral coefficients $\upsilon_k(\bm{s}),$ i.e. the first 9 principal components of the spectra obtained at location $\bm{s}$. This model allows us to link the two types of observation available from RaCA.

{The last model we consider uses the VNIR spectra to model the mean while considering spatial interactions with land-use for both the mean and the variance,}

\textbf{Model 6}

\begin{align*}
    y(\bm{s}) &=\gamma_{\ell(\bm{s})} + \sum_{k=1}^K\upsilon_k(\bm{s})\chi_k  + \bm{x}'(\bm{s}) \bm{\beta}_1 + \xi(\bm{s}) + \mu_{\ell}(\bm{s}) + \epsilon(\bm{s}), \; \bm{s} \in \mathcal{S}_{lab} \\
    \xi(\bm{s}) &= \sum_{j=1}^J \phi_j(\bm{s}) \eta_j \\
    \mu_{\ell}(\bm{s}) &= \sum_{j=1}^J \phi_j(\bm{s}) \kappa_{\ell j} \\
    \epsilon(\bm{s}) &\stackrel{ind}{\sim} N(0, \tau^2(\bm{s})) \\
    -\log\left(\tau^2(\bm{s})\right) &=\zeta_{\ell(\bm{s})} +  \bm{x}'(\bm{s}) \bm{\beta}_2 + \delta(\bm{s}) + \lambda_{\ell}(\bm{s}) \\
    \delta(\bm{s}) &= \sum_{j=1}^J \phi_j(\bm{s}) \alpha_j \\ 
    \lambda_{\ell}(\bm{s}) &= \sum_{j=1}^J \phi_j(\bm{s}) \nu_{\ell j}.
\end{align*} Model 6 is identical to model 4, however, now the mean is a linear function of the spectral coefficients $\upsilon_k(\bm{s}).$

\subsection{Cross-Validation Results}

We use five-fold cross validation to compare the six models. All models were run using Gibbs sampling for 5,000 iterations, and discarding the first 1,000 iterations as burn-in. Convergence was assessed through visual inspection of the traceplots, where no lack of convergence was detected. For each model, we compute the prediction mean squared error (MSE),
$$
\frac{1}{N} \sum_i \left(y_i(\bm{s}) - \widehat{y_i(\bm{s})}\right)^2,
$$
the mean squared error of variance (MSEV),
$$
\frac{1}{N} \sum_i \left\{\left(y_i(\bm{s}) - \widehat{y_i(\bm{s})}\right)^2 - \widehat{\tau^2_i(\bm{s})}\right\}^2,
$$
the 95\% prediction interval coverage rate (CR), the energy score (ES), and the average interval score (IS) \citep{gneiting2007strictly}. {Note that for this cross-validation study, in models 5 and 6 we assume that the spectral coefficients, $\upsilon_k(\bm{s}), k=1,\ldots,K$ are known for the out of sample locations. We will explore an approach to further predict these coefficients in Section~\ref{sec: analysis}.}  

A summary of the cross-validation results are presented in Table~\ref{Table: tab1}. We observe that the inclusion of the data from the spectra induces considerable improvement in the all the scores, with the exception of interval coverage. { In addition to this, we see that the incorporation of spatial dependence structure for both the mean and the variance results in superior predictive performance. Furthermore, the interaction between spatial dependence and land-use seems warranted, particularly in the mean structure. Models 5 and 6 are quite comparable, with no clear standout. Thus, appealing to parsimony, we choose model 5 as our working model moving forward.} 

\begin{table}[H]
\centering
\begin{tabular}{||c c c c c c ||} 
 \hline
 Model & MSE & MSEV & CR & IS & ES \\ [0.5ex] 
 \hline\hline
 Model 1 & 0.466 & 1.52 & 90.1\% &  4.43 & 12.14  \\ 
 \hline
 Model 2 & 0.399 & 1.60 & 94.9\% & 3.32  & 11.20 \\
 \hline
 Model 3 & {0.341} & {1.21} & {93.9\%} & {3.15}  & {10.3} \\
 \hline
 Model 4 & {0.342} & {1.19} & \textbf{94.9\%} & {3.25}  & {11.21} \\
 \hline
 Model 5 & \textbf{0.225} & {0.62} & {92.5\%} & {2.68}  & \textbf{8.34} \\
 \hline
 Model 6 & {0.225}   & \textbf{0.60}  & 93.8\%  & \textbf{2.58}  &  8.53  \\
 \hline
\end{tabular}
\caption{Five-fold cross-validation results on lab-measured SOC data from the RaCA.}
\label{Table: tab1}
\end{table}

\section{U.S. Soil Carbon Mapping}\label{sec: analysis}

{Using model 5, prediction of soil carbon across the CONUS requires a land-use category for each location as well as values for the spectral coefficients at each location. Next, we describe how these values are acquired, and then present complete soil carbon mapping results.}

\subsection{Land-Use Category Across the CONUS}

\begin{figure}[H]
\centering
\includegraphics[width=150mm,height=80mm]{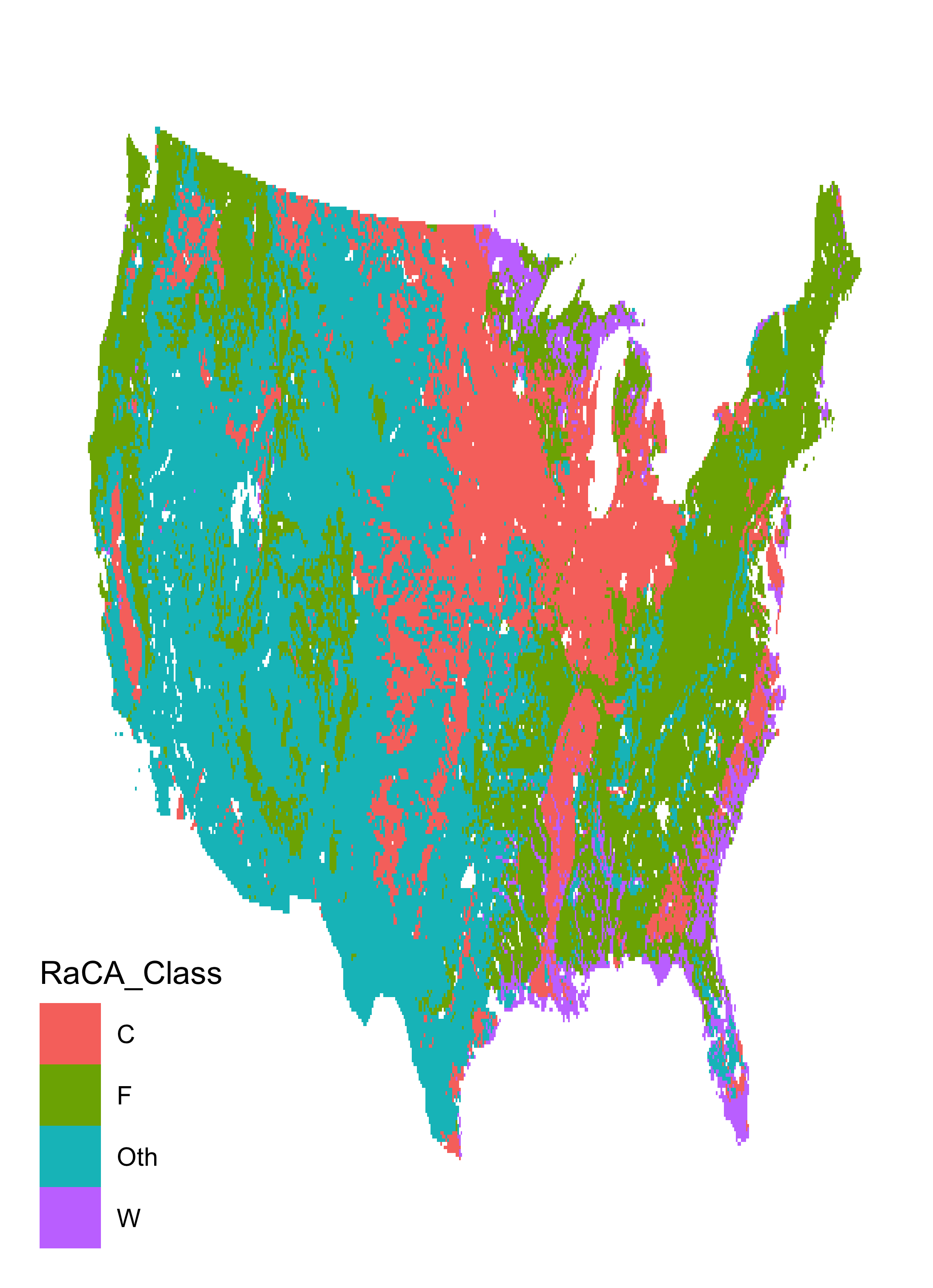}
\caption{Land-use category at the 10km scale for the CONUS. Land-use is derived from the 2013 National Land Cover Database. Land that is developed, barren, perennial ice, or open water is not shown.} 
\label{FIG: lulc}
\end{figure}

{\cite{wills2014overview} describe how the land-use categories reported by RaCA were derived from the United States Geological Survey 2013 National Land Cover Database (NLCD). The NLCD contains land-cover at an extremely fine resolution of 30 meters across the CONUS. Each location in the database identifies one of 16 land cover types. Each of these land cover types is associated with a specific land-use as described in \cite{wills2014overview}. These land-uses include cropland (C), forestland (F), wetland (W), and `other' (Oth), where `other' includes both pastureland and rangeland. In addition to this, there are many locations where the land-use category is not applicable, because the land is either developed, barren, perennial ice, or open water. We do not predict soil carbon at these locations. 

The NLCD is given at an extremely high resolution. We choose to reduce this resolution for two reasons. First, we wish to limit the computational overhead required to construct a map of soil carbon across the CONUS. Second, the geographic coordinates reported with the RaCA are truncated to two decimal places, or roughly 10 kilometers. Thus, we aggregate the NLCD to a 10 kilometer resolution by taking the modal land-use category. This 10 kilometer resolution represents the scale that we construct soil carbon predictions at. We denote these prediction locations as $\bm{s} \in \mathcal{P}.$ These land-use categories for each $\bm{s} \in \mathcal{P}$ are presented in Figure~\ref{FIG: lulc}. Here we see that the majority of cropland is in the Midwest, with modest sized regions in California and the Pacific northwest as well. Additionally, wetlands are primarily represented in the southeast and the Great Lakes region.}

\subsection{Predicting VNIR Spectra}

{As mentioned in Section~\ref{sec: cv}, we take Model 5 as our working model. However, prediction of soil carbon at spatial locations outside of our sample locations (i.e., $\bm{s} \in \mathcal{P}$) requires a prediction of the spectral coefficients, $\upsilon_k(\bm{s}), \; k=1,\ldots,K$ for these prediction locations. Importantly, the use of orthogonal principal components for the spectral basis expansion allows us to model each $\upsilon_k(\bm{s})$ independently.

A variety of methods to predict $\upsilon_k(\bm{s})$ for $\bm{s} \in \mathcal{P}$ were considered, including spatial models, as well as various nonlinear regression approaches. Ultimately, we found the dependence for these coefficients to be highly localized, with a K-nearest neighbors regression (K=1) leading to the most accurate predictions. That is, predictions were made using spectral coefficients from the nearest $\bm{s} \in \mathcal{S}.$ Again, the spatial coverage of these locations is given in Figure~\ref{FIG: all}.

Finally, we repeat the five-fold cross-validation exercise from Section~\ref{sec: cv} again, using the predicted spectral coefficients on held-out data points, rather than treating them as known. This resulted in an MSE and MSEV of 0.270 and 0.678 respectively. Although these are slightly higher than the values obtained when treating the spectral coefficients as known, they still present a substantial improvement over the models that did not incorporate the spectral information at all. The 95\% prediction interval coverage rate of 89.0\%  was slightly lower than Section~\ref{sec: cv}, likely due to the lack of uncertainty around the predicted spectral coefficients. However, the average interval score and energy score were 3.11 and 9.38 respectively, which again represent an improvement over the models that do not incorporate the spectra. Thus, even through we are predicting the spectral coefficients for out of sample locations, we have a strong indication that the resulting predictive distributions are characterizing SOC better than those from a heterogeneous spatial model that does not incorporate the VNIR spectra.
}
\subsection{SOC Across the CONUS}

{Using Model 5 alongside the land-use data provided by the NLCD, we predict SOC across the entire CONUS at a 10km resolution. The posterior mean predictions are shown in Figure~\ref{FIG: pred}. We see that the eastern US exhibits generally higher SOC content compared to other regions, followed closely by areas in the Great Lakes region, specifically northern Minnesota. This is likely driven by the proximity of wetlands in these areas. In contrast to this, the great plains region generally has the lowest SOC content in the CONUS. We can also see that SOC is heavily driven by land-use through comparison of the predictions in Figure~\ref{FIG: pred} to the land-use map given in Figure~\ref{FIG: lulc}. In particular, for the western US, there is a visible upward shift in SOC for forestland compared to other land-use categories in the vicinity.

\begin{figure}[H]
\centering
\includegraphics[width=150mm,height=80mm]{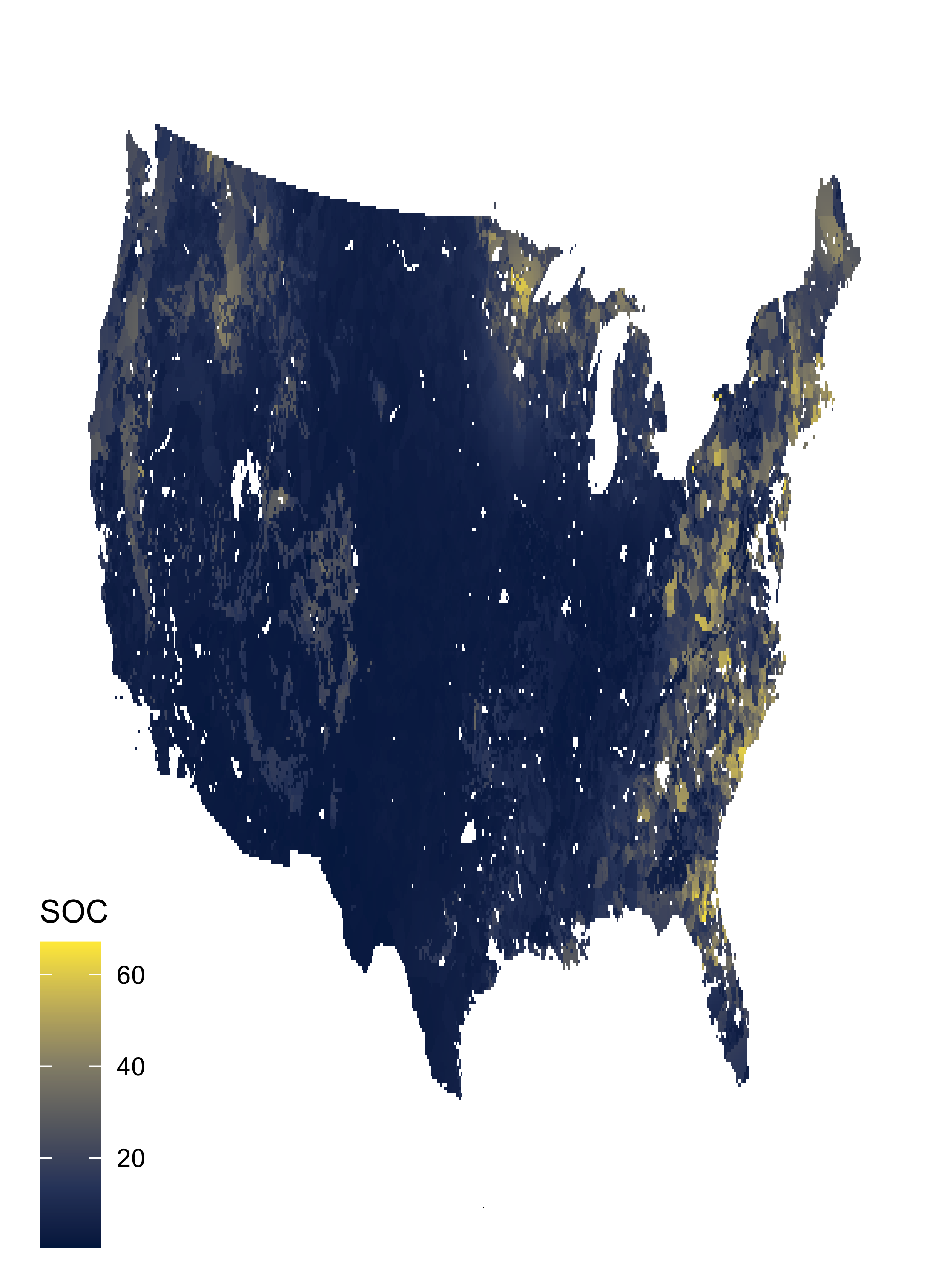}
\caption{Posterior mean of surface soil carbon prediction for the CONUS. Land that is developed, barren, perennial ice, or open water is not shown.} 
\label{FIG: pred}
\end{figure}

In addition to the posterior mean of predicted SOC, we show the posterior standard deviation (on the log scale) in Figure~\ref{FIG: se}. Similar to the mean predictions, the uncertainty around the predictions is linked with land-use category. For example, in the great plains region, there is a clear increase in uncertainty for cropland compared to other land uses in the same region.

\begin{figure}[H]
\centering
\includegraphics[width=150mm,height=80mm]{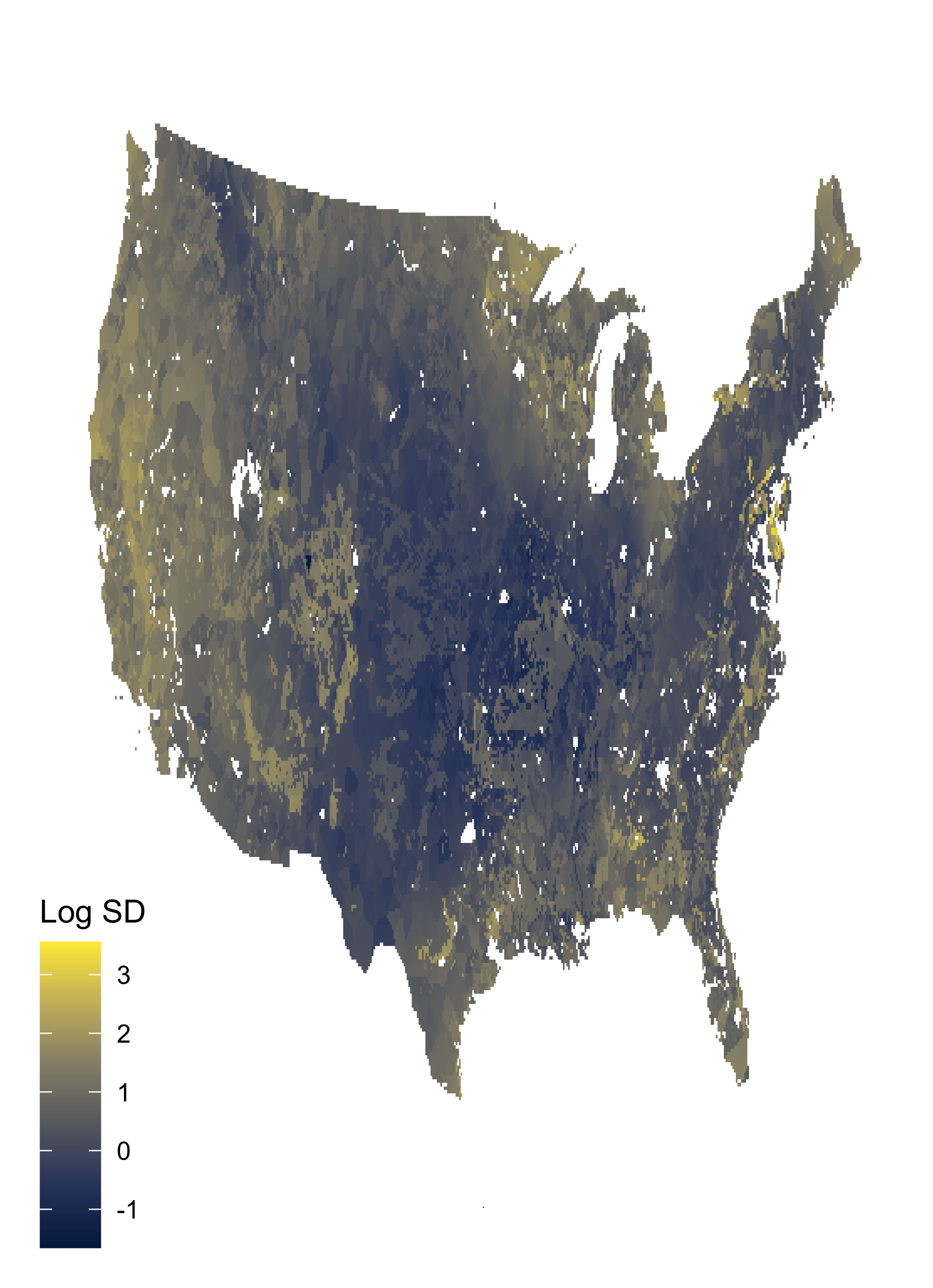}
\caption{Posterior standard deviation on the log scale of surface soil carbon prediction for the CONUS. Land that is developed, barren, perennial ice, or open water is not shown.} 
\label{FIG: se}
\end{figure}

\section{Discussion}\label{sec: discussion}

{Motivated by the goal of predicting soil organic carbon across the CONUS using the RaCA, we have developed a highly heterogeneous spatial model that incorporates spectral data. Importantly, our approach allows for the SOC mean function as well as the variance function to both vary across spatial location and by land-use category, a driving factor of SOC. Through a cross-validation study, we have shown that our selected model results in superior point and uncertainty estimation when compared to alternative models that either do not consider the VNIR spectra and/or do not have as flexible structure in the mean or variance.

Understanding the spatial distribution of soil carbon is an important component of determining soil quality and is closely related to the process of climate change. Yet, up until recently the task has been difficult due to the lack of a nationally representative dataset. The RaCA was able to help overcome this challenge by developing a large scale nationally representative data collection effort for soil carbon. One piece of this analysis that was not considered is the temporal dynamics of soil carbon. The RaCA was designed as a cross-sectional representation of soil carbon at a specific time. Soil carbon is a relatively slow evolving process, with time needed on the order of a decade typically to detect changes \citep{saby2008will}. Still, there is a time dynamic to SOC, and new large-scale datasets will need to be collected in the future to describe it. An interesting question remains of how to sample across both time and space in order to understand these dynamics.

{Along these lines, soil carbon at varying depths was not considered in this work. Most soil carbon is stored near the surface, motivating our approach of modeling soil surface carbon. However, as the interest for carbon sequestration continues to increase with climate change, exploring carbon below the surface level may be an important problem. Such analysis would require an extension of our approach that includes a third spatial dimension corresponding to soil depth.}

Another potential avenue related to this work that is worth exploring is the use of other types of spatially referenced covariates. For example, with constant advances in modern technology, there is increasing access to many types of satellite data. Some of these data might be correlated with SOC, such as weather or climate related data. The use of such covariates could further improve predictions of SOC and reduce uncertainty.

Ultimately the goal of this work was to produce a map of SOC across the CONUS with corresponding uncertainty quantification. We believe that this product is useful to soil scientists, but also those interested in carbon sequestration and its impact on climate change. The methods developed herein may also be more broadly useful to those working in other domains with highly heterogeneous data. Lastly, we hope that this work will help motivate future SOC data collection efforts, especially those with increased temporal availability.

}

\newpage
\bibliographystyle{apalike}

\begin{thebibliography}{}

\bibitem[Bradley et~al., 2018]{bradley2018computationally}
Bradley, J.~R., Holan, S.~H., and Wikle, C.~K. (2018).
\newblock Computationally efficient multivariate spatio-temporal models for
  high-dimensional count-valued data (with discussion).
\newblock {\em Bayesian Analysis}, 13(1):253--310.

\bibitem[Bradley et~al., 2019]{bradley2019bayesian}
Bradley, J.~R., Holan, S.~H., and Wikle, C.~K. (2019).
\newblock Bayesian hierarchical models with conjugate full-conditional
  distributions for dependent data from the natural exponential family.
\newblock {\em Journal of the American Statistical Association}, pages 1--16.

\bibitem[Brown et~al., 2001]{brown2001bayesian}
Brown, P.~J., Fearn, T., and Vannucci, M. (2001).
\newblock Bayesian wavelet regression on curves with application to a
  spectroscopic calibration problem.
\newblock {\em Journal of the American Statistical Association},
  96(454):398--408.

\bibitem[Brown et~al., 1994]{brown1994multivariate}
Brown, P.~J., Le, N.~D., and Zidek, J.~V. (1994).
\newblock Multivariate spatial interpolation and exposure to air pollutants.
\newblock {\em Canadian Journal of Statistics}, 22(4):489--509.

\bibitem[Chakraborty, 2012]{chakraborty2012bayesian}
Chakraborty, S. (2012).
\newblock Bayesian multiple response kernel regression model for high
  dimensional data and its practical applications in near infrared
  spectroscopy.
\newblock {\em Computational Statistics \& Data Analysis}, 56(9):2742--2755.

\bibitem[Gneiting and Raftery, 2007]{gneiting2007strictly}
Gneiting, T. and Raftery, A.~E. (2007).
\newblock Strictly proper scoring rules, prediction, and estimation.
\newblock {\em Journal of the American Statistical Association},
  102(477):359--378.

\bibitem[Gramacy and Lee, 2008]{gramacy2008bayesian}
Gramacy, R.~B. and Lee, H. K.~H. (2008).
\newblock Bayesian treed {G}aussian process models with an application to
  computer modeling.
\newblock {\em Journal of the American Statistical Association},
  103(483):1119--1130.

\bibitem[Grenier et~al., 2023]{GreSanMatt2023}
Grenier, I., Sans\'o, B., and Matthews, J.~L. (2023).
\newblock Multivariate nearest-neighbors {G}aussian processes with random
  covariance matrices.
\newblock Technical report, UCSC-SOE-23-01.

\bibitem[Guti{\'e}rrez et~al., 2014]{gutierrez2014bayesian}
Guti{\'e}rrez, L., Guti{\'e}rrez-Pe{\~n}a, E., and Mena, R.~H. (2014).
\newblock Bayesian nonparametric classification for spectroscopy data.
\newblock {\em Computational Statistics \& Data Analysis}, 78:56--68.

\bibitem[Higdon, 1998]{higdon1998process}
Higdon, D. (1998).
\newblock A process-convolution approach to modelling temperatures in the
  {N}orth {A}tlantic {O}cean.
\newblock {\em Environmental and Ecological Statistics}, 5:173--190.

\bibitem[Kim et~al., 2005]{kim2005analyzing}
Kim, H.-M., Mallick, B.~K., and Holmes, C.~C. (2005).
\newblock Analyzing nonstationary spatial data using piecewise {G}aussian
  processes.
\newblock {\em Journal of the American Statistical Association},
  100(470):653--668.

\bibitem[Kirsner and Sans{\'o}, 2020]{kirsner2020multi}
Kirsner, D. and Sans{\'o}, B. (2020).
\newblock Multi-scale shotgun stochastic search for large spatial datasets.
\newblock {\em Computational Statistics \& Data Analysis}, 146:106931.

\bibitem[Lemos and Sans{\'o}, 2009]{lemos2009spatio}
Lemos, R.~T. and Sans{\'o}, B. (2009).
\newblock A spatio-temporal model for mean, anomaly, and trend fields of
  {N}orth {A}tlantic sea surface temperature.
\newblock {\em Journal of the American Statistical Association},
  104(485):5--18.

\bibitem[Nunes et~al., 2021]{nunes2021soil}
Nunes, M.~R., Veum, K.~S., Parker, P.~A., Holan, S.~H., Karlen, D.~L., Amsili,
  J.~P., van Es, H.~M., Wills, S.~A., Seybold, C.~A., and Moorman, T.~B.
  (2021).
\newblock The soil health assessment protocol and evaluation applied to soil
  organic carbon.
\newblock {\em Soil Science Society of America Journal}, 85(4):1196--1213.

\bibitem[Parker et~al., 2021]{parker2021general}
Parker, P.~A., Holan, S.~H., and Wills, S.~A. (2021).
\newblock A general {B}ayesian model for heteroskedastic data with fully
  conjugate full-conditional distributions.
\newblock {\em Journal of Statistical Computation and Simulation},
  91(15):3207--3227.

\bibitem[Risser et~al., 2019]{risser2019nonstationary}
Risser, M.~D., Calder, C.~A., Berrocal, V.~J., and Berrett, C. (2019).
\newblock Nonstationary spatial prediction of soil organic carbon.
\newblock {\em The Annals of Applied Statistics}, 13(1):165--188.

\bibitem[Saby et~al., 2008]{saby2008will}
Saby, N.~P., Bellamy, P.~H., Morvan, X., Arrouays, D., Jones, R.~J., Verheijen,
  F.~G., Kibblewhite, M.~G., Verdoodt, A., {\"U}veges, J.~B., Freudenschuss,
  A., et~al. (2008).
\newblock Will {E}uropean soil-monitoring networks be able to detect changes in
  topsoil organic carbon content?
\newblock {\em Global Change Biology}, 14(10):2432--2442.

\bibitem[Sampson and Guttorp, 1992]{sampson1992nonparametric}
Sampson, P.~D. and Guttorp, P. (1992).
\newblock Nonparametric estimation of nonstationary spatial covariance
  structure.
\newblock {\em Journal of the American Statistical Association},
  87(417):108--119.

\bibitem[Schmidt and Guttorp, 2020]{schmidt2020flexible}
Schmidt, A.~M. and Guttorp, P. (2020).
\newblock Flexible spatial covariance functions.
\newblock {\em Spatial Statistics}, 37:100416.

\bibitem[Schmidt and O'Hagan, 2003]{schmidt2003bayesian}
Schmidt, A.~M. and O'Hagan, A. (2003).
\newblock Bayesian inference for non-stationary spatial covariance structure
  via spatial deformations.
\newblock {\em Journal of the Royal Statistical Society Series B: Statistical
  Methodology}, 65(3):743--758.

\bibitem[Smith et~al., 2020]{smith2020measure}
Smith, P., Soussana, J.-F., Angers, D., Schipper, L., Chenu, C., Rasse, D.~P.,
  Batjes, N.~H., van Egmond, F., McNeill, S., and Kuhnert, M. (2020).
\newblock How to measure, report and verify soil carbon change to realize the
  potential of soil carbon sequestration for atmospheric greenhouse gas
  removal.
\newblock {\em Global Change Biology}, 26(1):219--241.

\bibitem[Stingo et~al., 2012]{stingo2012bayesian}
Stingo, F.~C., Vannucci, M., and Downey, G. (2012).
\newblock Bayesian wavelet-based curve classification via discriminant analysis
  with markov random tree priors.
\newblock {\em Statistica Sinica}, 22(2):465.

\bibitem[Wijewardane et~al., 2016]{wijewardane2016prediction}
Wijewardane, N.~K., Ge, Y., Wills, S., and Loecke, T. (2016).
\newblock Prediction of soil carbon in the conterminous {U}nited {S}tates:
  {V}isible and near infrared reflectance spectroscopy analysis of the rapid
  carbon assessment project.
\newblock {\em Soil Science Society of America Journal}, 80(4):973--982.

\bibitem[Wills et~al., 2014]{wills2014overview}
Wills, S., Loecke, T., Sequeira, C., Teachman, G., Grunwald, S., and West,
  L.~T. (2014).
\newblock Overview of the {US} rapid carbon assessment project: {S}ampling
  design, initial summary and uncertainty estimates.
\newblock {\em Soil carbon}, pages 95--104.

\bibitem[Yang et~al., 2015]{yang2015bayesian}
Yang, W.-H., Wikle, C.~K., Holan, S.~H., Myers, D.~B., and Sudduth, K.~A.
  (2015).
\newblock Bayesian analysis of spatially-dependent functional responses with
  spatially-dependent multi-dimensional functional predictors.
\newblock {\em Statistica Sinica}, pages 205--223.

\bibitem[Zammit-Mangion et~al., 2022]{zammit2022deep}
Zammit-Mangion, A., Ng, T. L.~J., Vu, Q., and Filippone, M. (2022).
\newblock Deep compositional spatial models.
\newblock {\em Journal of the American Statistical Association},
  117(540):1787--1808.

\end{thebibliography}

\section*{Appendix A: Full Conditional Distributions}

Let $\bm{s}_1,\ldots,\bm{s}_n \in \mathcal{S}_{lab}$ and $\bm{\Omega}_y=\hbox{Diag}\left(1/\sigma^2(\bm{s}_1),\ldots,1/\sigma^2(\bm{s}_n)\right)$. Also let $\bm{y}=(y(\bm{s}_1),\ldots,y(\bm{s}_n))'$ and $\bm{\mu}_y=(\mu(\bm{s}_1),\ldots,\mu(\bm{s}_n))'$. Finally, let $\bm{X}_1$ be the $n \times p_1$ matrix with $i$th row equal to $\bm{x}_1(\bm{s}_i)$, $\bm{X}_2$ be the $n \times p_2$ matrix with $i$th row equal to $\bm{x}_2(\bm{s}_i)$, $\bm{\Psi}_1$ be the $n \times r_1$ matrix with $i$th row equal to $\bm{\psi}_1(\bm{s}_i)$, and $\bm{\Psi}_2$ be the $n \times r_2$ matrix with $i$th row equal to $\bm{\psi}_2(\bm{s}_i)$. Then the posterior distribution of the general model outlined in Section~\ref{sec: methods} can be sampled from using Gibbs sampling by iteratively sampling from the below full conditional distributions.

\begin{align*}
    \bm{\beta}_1 | \cdot & \propto \hbox{exp}\left(-\frac{1}{2}(\bm{y} - \bm{X}_1 \bm{\beta}_1 - \bm{\Psi}_1 \bm{\eta}_1)'\bm{\Omega}_y(\bm{y} - \bm{X}_1 \bm{\beta}_1 - \bm{\Psi}_1 \bm{\eta}_1) \right) \\
    & \times \hbox{exp}\left(-\frac{1}{2\sigma^2_{\beta_1}} \bm{\beta}'\bm{\beta} \right) \\
    \bm{\beta}_1 | \cdot & \sim \hbox{N}_{p_1}\left( 
        \bm{\mu}=(\bm{X}_1'\bm{\Omega}_y\bm{X}_1 + \frac{1}{\sigma^2_{\beta_1}} \bm{I}_{p_1})^{-1} \bm{X}_1'\bm{\Omega}_y(\bm{y} - \bm{\Psi}_1\bm{\eta}_1),
        \bm{\Sigma} = (\bm{X}_1'\bm{\Omega}_y\bm{X}_1 + \frac{1}{\sigma^2_{\beta_1}} \bm{I}_{p_1})^{-1}
    \right) 
\end{align*}
    
\begin{align*}
    \bm{\eta}_1 | \cdot & \propto \hbox{exp}\left(-\frac{1}{2}(\bm{y} - \bm{X}_1 \bm{\beta}_1 - \bm{\Psi}_1 \bm{\eta}_1)'\bm{\Omega}_y(\bm{y} - \bm{X}_1 \bm{\beta}_1 - \bm{\Psi}_1 \bm{\eta}_1) \right) \\
    & \times \hbox{exp}\left(-\frac{1}{2\sigma^2_{\eta_1}} \bm{\eta}'\bm{\eta} \right) \\
    \bm{\eta}_1 | \cdot & \sim \hbox{N}_{r_1}\left( 
        \bm{\mu}=(\bm{\Psi}_1'\bm{\Omega}_y\bm{\Psi}_1 + \frac{1}{\sigma^2_{\eta_1}} \bm{I}_{r_1})^{-1} \bm{\Psi}_1'\bm{\Omega}_y(\bm{y} - \bm{X}_1\bm{\beta}_1),
        \bm{\Sigma} = (\bm{\Psi}_1'\bm{\Omega}_y\bm{\Psi}_1 + \frac{1}{\sigma^2_{\eta_1}} \bm{I}_{r_1})^{-1}
    \right) 
\end{align*}

\begin{align*}
    \bm{\beta}_2 | \cdot & \propto \prod_{i=1}^n \hbox{exp} \left\{\frac{1}{2}\bm{x}_{2}(\bm{s}_i)'\bm{\beta}_2 - \frac{1}{2}\left(y(\bm{s}_i) - \mu(\bm{s}_i)\right)^2\hbox{exp}\left(\bm{\psi}_{2}(\bm{s}_i)'\bm{\eta}_2\right)\hbox{exp}\left(\bm{x}_{2}(\bm{s}_i)'\bm{\beta}_2\right) \right\} \\
    & \times \hbox{exp}\left\{\alpha \bm{1}_{p_2}' \alpha^{-1/2} \frac{1}{\sigma_{\beta_2}} \bm{I}_{p_2} \bm{\beta}_2 - \alpha \bm{1}_{p_2}' \hbox{exp}\left(\alpha^{-1/2} \frac{1}{\sigma_{\beta_2}} \bm{I}_{p_2} \bm{\beta}_2\right)  \right\} \\
            &= \hbox{exp}\left\{\bm{\alpha}_{\beta_2}' \bm{H}_{\beta_2} \bm{\beta}_2 - \bm{\kappa}_{\beta_2}' \hbox{exp}(\bm{H}_{\beta_2} \bm{\beta}_2)\right\} \\
        & \bm{H}_{\beta_2} = \left[
                            \begin{array}{c}
                            \bm{X}_2  \\
                            \alpha^{-1/2} \frac{1}{\sigma_{\beta_2}} \bm{I}_{p_2} 
                            \end{array}
                            \right], \quad 
        \bm{\alpha}_{\beta_2} = \left(\frac{1}{2} \bm{1}_n', \alpha \bm{1}_{p_2}'\right)', \quad \\
        & \bm{\kappa}_{\beta_2} = \left(\frac{1}{2}\left\{(\bm{y}-\bm{\mu}_y)^2 \odot \hbox{exp}(\bm{\Psi}_2 \bm{\eta}_2)\right\}', \alpha \bm{1}_{p_2}'\right)' \\
        \bm{\beta}_2 | \cdot &\sim \hbox{cMLG}(\bm{H}_{\beta_2}, \bm{\alpha}_{\beta_2}, \bm{\kappa}_{\beta_2})
\end{align*}

\begin{align*}
    \bm{\eta}_2 | \cdot & \propto \prod_{i=1}^n \hbox{exp} \left\{\frac{1}{2}\bm{\psi}_{2}(\bm{s}_i)'\bm{\eta}_2 - \frac{1}{2}(y(\bm{s}_i) - \mu(\bm{s}_i))^2\hbox{exp}(\bm{x}_{2}(\bm{s}_i)'\bm{\beta}_2)\hbox{exp}(\bm{\psi}_{2}(\bm{s}_i)'\bm{\eta}_2) \right\} \\
    & \times \hbox{exp}\left\{\alpha \bm{1}_{r_2}' \alpha^{-1/2} \frac{1}{\sigma_{\eta_2}} \bm{I}_{r_2} \bm{\eta}_2 - \alpha \bm{1}_{r_2}' \hbox{exp}\left(\alpha^{-1/2} \frac{1}{\sigma_{\eta_2}} \bm{I}_{r_2} \bm{\eta}_2\right)  \right\} \\
            &= \hbox{exp}\left\{\bm{\alpha}_{\eta_2}' \bm{H}_{\eta_2} \bm{\eta}_2 - \bm{\kappa}_{\eta_2}' \hbox{exp}(\bm{H}_{\eta_2} \bm{\eta}_2)\right\} \\
        & \bm{H}_{\eta_2} = \left[
                            \begin{array}{c}
                            \bm{\Psi}_2  \\
                            \alpha^{-1/2} \frac{1}{\sigma_{\eta_2}} \bm{I}_{r_2} 
                            \end{array}
                            \right], \quad 
        \bm{\alpha}_{\eta_2} = \left(\frac{1}{2} \bm{1}_n', \alpha \bm{1}_{r_2}'\right)', \quad \\
        & \bm{\kappa}_{\eta_2} = \left(\frac{1}{2}\left\{(\bm{y}-\bm{\mu}_y)^2 \odot \hbox{exp}(\bm{X}_2 \bm{\beta}_2)\right\}', \alpha \bm{1}_{r_2}'\right)' \\
        \bm{\eta}_2 | \cdot &\sim \hbox{cMLG}(\bm{H}_{\eta_2}, \bm{\alpha}_{\eta_2}, \bm{\kappa}_{\eta_2})
\end{align*}

\begin{align*}
    \sigma^2_{\eta_1} | \cdot & \propto \left(\sigma^2_{\eta_1}\right)^{-r_1/2} \hbox{exp}\left(-\frac{1}{2 \sigma^2_{\eta_1}} \bm{\eta}_1'\bm{\eta}_1 \right) \times \left(\sigma^2_{\eta_2}\right)^{-a -1}\hbox{exp}\left(-\frac{b}{\sigma^2_{\eta_1}}\right) \\ 
    &= \left(\sigma^2_{\eta_1}\right)^{-(a + r_1/2) - 1} \hbox{exp}\left\{-\frac{1}{\sigma^2_{\eta_2}}\left(b + \frac{\bm{\eta}_1'\bm{\eta}_1}{2}\right) \right\} \\
    \sigma^2_{\eta_1} | \cdot & \sim \hbox{IG}\left(a + \frac{r_1}{2}, \; b + \frac{\bm{\eta}_1'\bm{\eta}_1}{2} \right)
\end{align*}

\begin{align*}
    \frac{1}{\sigma_{\eta_2}} | \cdot & \propto \hbox{exp}\left\{\alpha \bm{1}_{r_2}' \alpha^{-1/2} \frac{1}{\sigma_{\eta_2}} \bm{I}_{r_2} \bm{\eta}_2 - \alpha \bm{1}_{r_2}' \hbox{exp}\left(\alpha^{-1/2} \frac{1}{\sigma_{\eta_2}} \bm{I}_{r_2} \bm{\eta}_2\right)  \right\} \\
        & \quad \times \hbox{exp}\left\{\omega \frac{1}{\sigma_{\eta_2}} - \rho \, \hbox{exp}\left(\frac{1}{\sigma_{\eta_2}}\right)\right\} \times I(\sigma_{\eta_2} > 0) \\
                & = \hbox{exp} \left\{\bm{\omega}_{\sigma}' \bm{H}_{\sigma} \frac{1}{\sigma_{\eta_2}} - \bm{\rho}_{\sigma}' \hbox{exp} \left( \bm{H}_{\sigma} \frac{1}{\sigma_{\eta_2}} \right) \right\} \times I(\sigma_{\eta_2} > 0) \\
        & \bm{H}_{\sigma} = (\alpha^{-1/2} \bm{\eta}_2', 1)' \quad
        \bm{\omega}_{\sigma} = (\alpha \bm{1}_{r_2}', \omega)' \quad 
        \bm{\rho}_{\sigma} = (\alpha \bm{1}_{r_2}', \rho)'\\
        \frac{1}{\sigma_{\eta_2}} | \cdot & \sim \hbox{cMLG}(\bm{H}_{\sigma}, \bm{\omega}_{\sigma}, \bm{\rho}_{\sigma}) \times I(\sigma_{\eta_2} > 0) 
\end{align*}

\end{document}